\documentclass[cameraready]{Interspeech}

\usepackage[utf8]{inputenc}
\usepackage[T1]{fontenc}
\usepackage{CJKutf8}

\title{Automatic Speech Recognition for Multilingual Oral History Research}

\author[affiliation={1,2,3}, orcid=0000-0002-8483-0540, equalcontribution, correspondingauthor]{Sidney}{Wong}
\author[affiliation={3}, equalcontribution]{Chelsea}{Wong She}
\author[affiliation={3}, equalcontribution]{Eda}{Tang}
\author[affiliation={3}, equalcontribution]{Tiana}{Marshall Wong}
\author[affiliation={3}]{Debbie}{Sew Hoy}
\author[affiliation={3}]{Chelsea}{Wong}

\address{
    $^1$ Centre of Sustainability Research, University of Otago, New Zealand \\
    $^2$ Te Pūnaha Matatini Centre of Research Excellence for Complex Systems, New Zealand \\
    $^3$ New Zealand Chinese Association, New Zealand 
}

\email{sidney.wong@otago.ac.nz}


\usepackage{comment}

\begin{document}

\maketitle

\begin{abstract}

    This paper offers a unique perspective on how speech technologies are being adopted by community-led heritage language preservation and revitalisation initiatives. As a community-led language maintenance strategy, oral histories play a crucial role in Cantonese language revitalisation in New Zealand. The development of Automatic Speech Recognition (ASR) toolkits, such as Whisper, have expedited what has often been a resource and time-intensive process of transcribing oral history collections. However, there is limited research into the effectiveness of ASR toolkits when applied to code-switched language contexts. Based on Word Error Rate (WER), the best performing Whisper model configuration achieved a WER of 12.10 at the expense of accurately transcribing unsupported non-English segments. However, Whisper remains a useful tool by providing a first-pass transcription using only 1\% of the estimated time otherwise needed for manual transcription.
\end{abstract}

\section{Introduction}

    With 54,417 speakers as of the 2023 Census \cite{stats_nz_2023_2024}, Cantonese is the seventh most spoken language in New Zealand and makes up one of the largest heritage language communities in the country. Cantonese-speaking Chinese communities have had a established presence in New Zealand since the 1840s \cite{ng_ninety_1962}. Despite being one of the largest established heritage language communities in the country, Cantonese-speaking communities in New Zealand have experienced significant levels of language shift and loss \cite{chen_chinese_2018}. A survey of Cantonese-speaking families living in the Wellington in the 1990s found that over the course of three generations, 70.96\% of people in Cantonese-speaking families no longer spoke Cantonese \cite{holmes_language_1993}. This level of language shift surpassed the rates observed in other heritage language speaking families in the same period.
    
    Researchers have attributed the process of language shift due to sustained legislative discrimination such as the introduction of the Chinese Poll Tax in 1881 \cite{wang_chinese_2023}\cite{holmes_language_1993}. More direct efforts to suppress the transmission of Cantonese resulted in a complete ban of language teachers from entering New Zealand \cite{ip_new_2005}\cite{ip_chinese_2013}. Other external factors such as social exclusion and demands for assimilation from outside the community \cite{yee_coping_2003}. Grass root attempts to curtail this trend has seen a rise in oral histories aim to document the history \textit{from below} \cite{lynd_oral_1993}. Unlike traditional histories, these oral histories are co-created between interviewers and interviewees and those who provide the narratives remain in control of the narratives produced during the interviews. Oral histories involving the New Zealand Chinese communities have largely focussed on the intersections of their lived experiences and the wider community \cite{tung_jung_association_of_new_zealand_inc_tung_2003}\cite{thorpe_otaki_2004}\cite{chang_oral_2005}\cite{gee_guangdong_2012}. 
    
    Oral histories and other artefacts (such as folk songs, proverbs, and storytelling) play an important role in the preservation and maintenance of linguistic and cultural knowledge \cite{ramesh_role_2025}. This is evident in the relative success of language revitalisation efforts of te reo Māori - the indigenous language of New Zealand - in leveraging language technologies \cite{james_developing_2020}. As heritage language policy and planning rely on community members \cite{chen_micro_2024}, the success of Cantonese revitalisation remains uncertain without legislative support. Both translating and transcribing oral histories remain a challenge in producing these linguistic and cultural artefacts. Simply put, there remains a barrier in accessing the oral histories of Cantonese-speaking elders further proliferating the impacts of language shift. Therefore, the use of open-source and accurate language technologies is particularly important for often low and under resourced language communities \cite{foley_building_2018}. In recognition of these resource constraints presented to heritage language communities, we ask how suitable are open-source speech recognition models when applied to multilingual oral history contexts?

\section{Related Works}

    In New Zealand, oral historians tend to place more emphasis on a time-coded abstracts of the interviews for the purposes of providing an index of named entities such as people and place names mentioned throughout a recording, rather than transcripts \cite{manatu_taonga__ministry_for_culture_and_heritage_processing_2024}. While the recording of the interview should be viewed as the primary source, a written transcript becomes a research tool and preservation format \cite{bergen_transcribing_2019}. Transcripts, when processed with careful editorial considerations, can be especially useful tools for analysing and presenting oral histories, which in turn can be impactful resources for histories that otherwise remain undocumented \cite{ritchie_doing_2015}. There is also the issue of practicality in transcribing hours of audio recordings. A survey of 51 linguists found that one hour of recordings may take up to 40 hours to produce word-level transcriptions \cite{foley_building_2018}. This is untenable for under-resourced language communities including minority and heritage language communities. 
    
    Automatic speech recognition (ASR) technologies have been readily adopted by researchers to support language documentation and revitalisation for indigenous and heritage language communities \cite{james_developing_2020}. Counterintuitively, the primary aim of ASR in under-resourced language conditions is to produce a first-pass transcription that is then corrected and edited by speakers \cite{prudhommeaux_automatic_2021}. As a form of speech processing, ASR transforms speech to sequence of words. Early ASR systems relied on statistical language models \cite{huang_overview_2010} and gave rise to open-source ASR toolkits such as KALDI \cite{povey_kaldi_2011}, which later adopted deep neural network (DNN) framework architectures. These toolkits can be integrated into browser-based software to support oral historians during the transcription process \cite{draxler_speech_2024}. As an example, ASR was extensively used to transcribe 18 hours of multilingual spoken narratives from the \textit{Voices from Ravensbrück} in English, Dutch, and German \cite{calamai_voices_2021}\cite{draxler_transcribing_2025}. 
    
    The introduction of the encoder-decoder transformer architecture accelerated the development of neural-network based ASR systems \cite{vaswani_attention_2017}. Unlike statistical language models, transformer-based ASR systems treat speech-to-text as a sequence-to-sequence task. The Whisper toolkit is now considered state-of-the-art ASR system trained on 680,000 hours of audio recordings \cite{radford_robust_2023}. Oral historians have promoted Whisper as a viable tool for transcription \cite{draxler_speech_2024}. While Whisper has been effectively applied to Cantonese ASR \cite{an_cantonese_2025}\cite{zhang_lora-int8_2025}, Cantonese transcription is not simply a case of speech-to-text. Cantonese exists in a state of diglossia \cite{cheang_diglossia_2022}, where the written standard is derived from Standard Written Chinese based on Mandarin. While a written form of Cantonese exists, it is rarely taught in formal educational contexts. Furthermore, linguistic variation exists within the topolects of Cantonese as shown in Table \ref{tab:topolect_compare} \cite{norman_chinese_1988}. Similarly, there is limited evidence to suggest these toolkits are suitable for code-switched language contexts \cite{zhao_adapting_2025}. 

    \begin{table}[h]
    
        \centering

        \caption{Phonetic variation across Cantonese topolects.}
        \label{tab:topolect_compare}
        
        \begin{tabular}{lcccc}
        
            \toprule
            \textbf{Topolect} & \textbf{`I, me'} & \textbf{`you'} & \textbf{`he/she'} & \textbf{`we'} \\
            \midrule
            Guangzhou & \textipa{\ng}o\tone{24} & nei\tone{24} & k\super{h}\textipa{\o}y\tone{24} & \textipa{\ng}o\tone{24}tei\tone{33} \\
            Zhongshan & \textipa{\ng}o\tone{13} & ni\tone{13} & k\super{h}y\tone{51} & \textipa{\ng}o\tone{13}ti\tone{22} \\
            Yangjiang & \textipa{\ng}o\tone{21} & nei\tone{21} & kei\tone{443} & \textipa{\ng}ok\tone{24} \\
            Taishan & \textipa{\ng}o\tone{33} & ni\tone{33} & k\super{h}ui\tone{33} & \textipa{\ng}oi \\
            Tengxian & \textipa{\ng}\textopeno\tone{24} & ni\tone{24} & ky\tone{24} & \textipa{\ng}\textopeno\tone{24}ti\tone{22} \\

            \bottomrule
        \end{tabular}
        
    \end{table}

\section{Methodology}

    The purpose of this paper is to address a gap in the literature to understand not only \textit{why} oral historians should use ASR, but also to address \textit{how} oral historians could use ASR toolkits. More importantly, we highlight the limitations of ASR in an applied real-world context. We created a corpus of oral history interviews with community elders who were proficient in English and Cantonese (including related varieties such as Taishanese). The research team consisted of members of the New Zealand Chinese communities including heritage language speakers of Cantonese. In this section, we provide an overview of the corpus dimensions and data collection procedures. We then describe our speech processing procedures.

\subsection{Data Collection and Corpus Dimensions}

    The interviews were carried out by two members of the research team who are fluent speakers of English and variable proficiency of Cantonese. Project members received oral history training and all attempts have been made to comply with existing best practice approaches in oral history research.
    
\subsubsection{Speaker Characteristics}

    The corpus included five speakers who were recruited through the personal networks of the research team. We restricted the selection criteria included first and second generation New Zealand Cantonese aged 65 years and over who were born or migrated to New Zealand before 1980. This period was chosen as it aligned with legislative changes which have had a significant impact on New Zealand Chinese communities \cite{ng_ninety_1962}. The corpus consisted of two female and three male speakers.

\subsubsection{Recording Procedures}
    
    The interviews were recorded using a Zoom H5 audio recorder. Prior to the recording sessions, prompts were provided to the speakers. The first 20-30 minutes of each interview were dedicated to demographic questions such as their place of birth (including ancestral affiliations) or their educational background. Interviewees were actively encouraged to use any language.

\subsubsection{Ethical Considerations and Data Availability}

    Speakers consented that the audio recordings could be used to develop resources such as podcasts, teaching resources, documentaries, and published books/articles. The audio recordings will made be available through a public archive. As some of the speakers include prominent members of the New Zealand Chinese community, for this reason, one speaker have elected to release the audio recordings posthumously. One speaker wanted their recordings completely available without the need to seek further permission. The audio recordings remain the intellectual property of the speakers.

    \begin{table*}[h]
    
        \centering

        \caption{Predicted outputs of a 2 second segment from reference set.}
        \label{tab:semantic_performance}
        
        \begin{tabular}{lll}
        
            \toprule
            \textbf{Size} & \textbf{Specified} & \textbf{Unspecified} \\
            \midrule
            \textsc{tiny} & did you ever sleep? Was, you know, did you see that? & do you ever see her? Or, you know, do you ever see her? \\
            \textsc{base} & Neil Thiela, or Neil Thiela, & Neil Thiela, or Neil Thiela, \\
            \textsc{small} & New Zealand or New Zealand, & Nihusila was, you know, Nihusila. \\
            \textsc{medium} & New Zealand or New Zealand, & Ni hui xi la, or, you know, ni hui xi la, \\
            \textsc{large-v1} & Niu Hsila was, you know, Niu Hsila. & - \\
            \textsc{large-v2} & Nehuseela was, you know, Nehuseela, & Ni hui hila. Or, you know, ni hui hila. \\
            \textsc{large-v3} & nǐ hù xī lè, or, you know, nǐ hù xī lè, & Nihusila was, you know, Nihusila. \\
            \textsc{turbo} & , you know, , & Nihusila, or, you know, Nihusila, \\
            \bottomrule
        \end{tabular}
        
    \end{table*}

\subsection{Speech Recognition}

    Our pilot corpus included 12 hours and 42 minutes of audio recordings. A rough estimate would put this at 508 hours to manually transcribe the entire corpus not including the time needed to translate between language conditions. As the state-of-the-art general purpose ASR model, we tested the suitability of Whisper to transcribe our multilingual oral history recordings \cite{radford_robust_2023}. We describe the data preprocessing, model training, and evaluation procedures below. We carried out our data manipulation, model training and evaluation in a Google Colab environment with Python 3 Google Compute Engine backend (GPU) enabled.

\subsubsection{Data Preprocessing}

    We installed \textsc{ffmpeg} to enable command-line manipulation of the audio files in the Google Colab environment. As dual microphones were used to record the speaker and the interviewer, we used the \textsc{pydub} Python package to overlay the recordings to produce one audio file.

\subsubsection{Pretrained Model}

    Included within the training data were 11,731 hours of audio for translation and 23,446 hours of audio for ASR in `Chinese'. Whisper models vary in size between 39 million (tiny) to 1,550 million (large) parameters. The six model sizes are: tiny (39 M), base (74 M), small (244 M), medium (769 M), large (1,550 M), and turbo (809 M). The large models are further broken down to large-v1, large-v2, and large-v3. In addition to the eight Whisper models, we included an English-specified and language-unspecified condition using Whisper's internal language detection algorithm to determine the primary language of the recording. This was based on an initial 30-second sample of the audio recording.

\subsubsection{Evaluation}
        
    The primary measure of model performance for ASR models is the Word Error Rate (WER). This measure is based on the number of words that differ between the reference and the prediction. Using the Common Voice 15 as a benchmark, the WER for English for the Whisper \texttt{large-v3} model was 9.3, for Hong Kong Cantonese was 15.9, and for Mainland Cantonese was 10.9 (as for Mainland Mandarin this was 12.8 and for Taiwan Mandarin this was 8.2). In addition to WER, we used the JiWER Python package to generate the following model performance metrics:
    
    \begin{itemize}
        \item Character Error Rate (CER): the number of characters that differ between the reference and the prediction.
        \item Match Error Rate (MER): the proportion of words that are misaligned between the reference and the prediction.
        \item Word Information Preserved (WIP): measures the proportion of information from the reference that is preserved in the prediction. The inverse of WIP is Word Information Lost (WIL).
    \end{itemize}

\subsubsection{Reference Set}

    We manually transcribed one audio recording of one male speaker that was 9 minute and 33 second long. The speaker consented to the use of their recordings for further analysis. Including the interviewer, the entire passage included 1,198 words code-switching between English, te reo Māori, and Chinese languages including Cantonese and Taishanese. The development of the reference set highlighted some key challenges for multilingual ASR. For example, English does not share the same orthographic conventions with Sinitic languages. For example, the primary Romanisation scheme for Cantonese is (\begin{CJK}{UTF8}{bsmi}粵拼\end{CJK}; `Linguistic Society of Hong Kong Cantonese Romanisation Scheme'), often referred to as Jyutping. However, Jyutping is generally restricted to academic or language learning contexts. For this reason, we have manually transcribed the non-English segments for all Sinitic languages (e.g., Cantonese, Taishanese, or Mandarin) with traditional Chinese characters. We used the \textit{tohutō} (`macron') for non-English segments in te reo Māori.

\section{Results}

    \begin{table}[h]
    
        \centering

        \caption{Performance metrics for English-specific models}
        \label{tab:english_specified}
        
        \begin{tabular}{lcccc}
        
            \toprule
            \textbf{Size} & \textbf{WER} & \textbf{CER} & \textbf{MER} & \textbf{WIP} \\
            \midrule
            \textsc{tiny} & 40.07 & 20.52 & 37.04 & 44.15 \\
            \textsc{base} & 29.47 & 16.14 & 28.04 & 58.11 \\
            \textsc{small} & 26.04 & 15.49 & 25.70 & 63.06 \\
            \textsc{medium} & 19.78 & 12.25 & 19.51 & 71.99 \\
            \textsc{large-v1} & 18.20 & \textbf{10.05} & \textbf{18.02} & 73.08 \\
            \textsc{large-v2} & \textbf{17.78} & 10.34 & 17.30 & \textbf{75.55} \\
            \textsc{large-v3} & 30.05 & 19.21 & 27.76 & 61.43 \\
            \textsc{turbo} & 24.62 & 11.18 & 23.64 & 61.94 \\
            \bottomrule
        \end{tabular}
        
    \end{table}

    \begin{table}[h]
    
        \centering

        \caption{Performance metrics for language-unspecified models}
        \label{tab:language_unspecified}
        
        \begin{tabular}{lccccc}
        
            \toprule
            \textbf{Size} & \textbf{WER} & \textbf{CER} & \textbf{MER} & \textbf{WIP} & \textbf{LAN} \\
            \midrule
            \textsc{tiny} & 36.39 & 21.38 & 34.71 & 49.06 & mri \\
            \textsc{base} & 29.47 & 16.14 & 28.04 & 58.11 & eng \\
            \textsc{small} & 40.65 & 27.89 & 40.08 & 45.47 & mri \\
            \textsc{medium} & 20.37 & 10.52 & 19.77 & 70.53 & mri \\
            \textsc{large-v1} & 24.37 & 13.62 & 24.25 & 64.58 & mri \\
            \textsc{large-v2} & 21.54 & 13.66 & 21.08 & 69.30 & mri \\
            \textsc{large-v3} & \textbf{12.10} & \textbf{8.73} & \textbf{11.89} & \textbf{84.39} & mri \\
            \textsc{turbo} & 24.62 & 11.18 & 23.64 & 61.94 & eng \\
            \bottomrule
        \end{tabular}
        
    \end{table}
    
    We now present the findings of our analysis. Using the manually transcribed transcript as the ground truth reference, we compared the model performance of seven Whisper models. We present the model performance metrics in the English-specific condition in Table \ref{tab:english_specified} and the language-unspecified condition in Table \ref{tab:language_unspecified}. Starting with the English-specific condition (Table \ref{tab:english_specified}), the model with the best performance was \textsc{large-v2} size based on an WER of 17.78\%; meanwhile, the model size with the worst performance was the \textsc{tiny} size with an WER 40.07\%. If we considered model performance across all metrics, then both \textsc{large-v1} and \textsc{large-v2} performed similarly.

    \begin{table*}[t]
      \caption{Reference and predicted output from male speaker.}
      \label{tab:whisper_output}
      \centering
      \begin{tabular}{cp{0.3\linewidth}p{0.3\linewidth}p{0.2\linewidth}}
        \toprule
        \textbf{Line} & \textbf{Reference} & \textbf{Prediction} & \textbf{Gloss} \\
        \midrule
        1 
        & Mum's from Hong Kong.
        & Mum's from Hong Kong. 
        & \\
        2
        & And so she was, you know, \underline{s\textsci{k}\textcorner\tone{22}f\textturna\textlengthmark{n}\tone{22}}. 
        & And so she was, you know, Sik Fan. 
        & s\textsci{k}\textcorner\tone{22}f\textturna\textlengthmark{n}\tone{22}: \begin{CJK}{UTF8}{bsmi}食飯\end{CJK} `to eat' \\
        3 
        & And my dad will be \underline{hiak\textcorner\tone{33}f\textturna{n}\tone{21}}.
        & And my dad will be Help Fan.
        & hiak\textcorner\tone{33}f\textturna{n}\tone{21}: \begin{CJK}{UTF8}{bsmi}吃飯\end{CJK} `to eat'\\
        4 
        & And so, you know, \underline{um} \underline{wun\tone{45}} and \underline{w\textturna{n}\tone{45}}.
        & And so, you know, Wan and Wan.
        & wun\tone{45}/w\textturna{n}\tone{45}: \begin{CJK}{UTF8}{bsmi}碗\end{CJK} `bowl'\\
        5 
        & \underline{v\textopeno{n}\tone{33}}, \underline{v\textopeno{n}\tone{33}}, \underline{v\textopeno{n}\tone{33}} is the \underline{bowl}.
        & Wan, Wan, Wan is the bull.
        & v\textopeno{n}\tone{33}: \begin{CJK}{UTF8}{bsmi}碗\end{CJK} `bowl'\\
        6 
        & And \underline{w\textturna{n}\tone{45}} is the other bowl in Chinese.
        & Wan, Wan is the other bull in Chinese.
        & w\textturna{n}\tone{45}: \begin{CJK}{UTF8}{bsmi}碗\end{CJK} `bowl' \\
        7 
        & So \underline{t\textopeno\textlengthmark\tone{55}t\texttoptiebar{s}\textepsilon\textlengthmark\tone{22}},
        & So Dou Jie,
        & t\textopeno\textlengthmark\tone{55}t\texttoptiebar{s}\textepsilon\textlengthmark\tone{22}: \begin{CJK}{UTF8}{bsmi}多謝\end{CJK} `thank you' \\
        8 
        & \underline{\textopeno\textlengthmark\tone{33}ti\textepsilon\tone{21}}.
        & or Dear.
        & \textopeno\textlengthmark\tone{33}ti\textepsilon\tone{21}: \begin{CJK}{UTF8}{bsmi}多謝\end{CJK} `thank you' \\
        9 
        & And so we sort of brought up both, you know, not that we knew any difference. We said, oh, yeah, okay. That is how it is, right? When you're not saying, oh, this isn't Cantonese
        & And so we sort of brought up both, you know, not that we knew any difference. We said, oh, yeah, okay. That is how it is, right? When you're not saying, oh, this isn't Cantonese
        & \\
        10 
        & this isn't \underline{\textbeltl{i}\tone{33}jip\textcorner\tone{55}}.
        & this isn't Siyip.
        & \textbeltl{i}\tone{33}jip\textcorner\tone{55}: \begin{CJK}{UTF8}{bsmi}四邑\end{CJK} `Taishanese'\\
        \bottomrule
      \end{tabular}
    \end{table*}

    In terms of the language-unspecified condition, the model with the best performance was the \textsc{large-v3} size with an WER of 12.10 and the worst performing model was the \textsc{small} size with an WER of 40.65. If we considered model performance across all metrics, then the \textsc{large-v1} size outperformed across the four metrics. When we inspected the results from Whisper's language detection algorithm, six of the eight model configurations detected te reo Māori as the primary language spoken (with the exception of \textsc{base} and \textsc{turbo} models which detected English as the primary language).

    Based on the performance metrics alone, the language-unspecified \textsc{large-v3} had the best model performance on the reference set. However, performance metrics like WER and CER do not necessarily suggest good performance across semantic or multilingual contexts. In order to do this, we must evaluate the predicted outputs impressionistically. We compared the predicted outputs with one two-second utterance from the reference set where we observed code-switching between Cantonese, English, and Taishanese (where we have provided the orthographic and literal translations in parentheses): 
    
    \begin{quote}
        
    nei\tone{24}h\textipa{\o}y\tone{24}sei\tone{24}la\textlengthmark\tone{53} (\begin{CJK}{UTF8}{bsmi}你去死啦\end{CJK}; `Go to hell'), or you know, ni\tone{33}hu\textlengthmark{y}\tone{11}\textbeltl{ei}\tone{11}la\textlengthmark\tone{33} (\begin{CJK}{UTF8}{bsmi}你去死啦\end{CJK}; `Go to hell')

    \end{quote}

    We present the predicted outputs in Table \ref{tab:semantic_performance}. Over half of the model transliterated the non-English segments in the predicted outputs. However, we also observed examples of hallucinations (e.g., `did you ever sleep? Was, you know, did you see that?' in \textsc{tiny} and `New Zealand or New Zealand' in \textsc{small} and \textsc{medium}) in the English-specified condition. With the exception of \textsc{tiny} and \textsc{large-v1}, the language-unspecified models all offered transliterations of the non-English segments. While we could not observe examples of Jyutping, the English-specified \textsc{large-v3} model and the language-unspecified \textsc{medium} model transliterated the non-English segments with a modified version of Hànyǔ Pīnyīn (\begin{CJK}{UTF8}{bsmi}漢語拼音\end{CJK}; `Chinese Phonetic Alphabet'). Hànyǔ Pīnyīn is the primary Romanisation system for Mandarin Chinese which suggested the models detected features associated with Mandarin Chinese. However, all models failed to distinguish phonetic differences between Cantonese and Taishanese segments using the same transliteration.

\section{Discussion}

    The best performing Whisper model configuration based on the reference set was the language-unspecified \textsc{large-v3}. In summary, the WER was 12.10, the CER was 8.73, the MER was 11.89, and the WIP was 84.39. In comparison to existing benchmarks, the WER fell within the existing range for English (9.3) and Hong Kong Cantonese (15.9) with the Common Voice 15 as benchmark \cite{radford_robust_2023}. These results suggest that Whisper is indeed a viable tool in transcribing oral history audio recordings \cite{draxler_speech_2024}. However, this assumes oral history recordings are conducted in one language by default. In the case of the reference set - which included English, Cantonese, Taishanese, and te reo Māori - some Whisper model configurations performed poorly in not only detecting non-English segments, but also transliterating non-English segments.
    
    In order to examine the downstream impacts of this issue, we included in an extended excerpt as shown in Table \ref{tab:whisper_output} where we compared the predicted output from the language-unspecified \textsc{large-v1} model with a 38 second segment from the same speaker used for the reference set. We included this segment as it shows how the boundaries between Cantonese and Taishanese are unclear even for the speaker. While the Whisper ASR model attempted to transliterate non-English segments in Cantonese with limited success (observed in Lines 2, 4, and 6), it struggled to transliterate non-English segments in Taishanese (observed in Lines 3, 5, 8, and 10). For the non-English segments in Taishanese, we observed hallucinations (`or Dear' in Line 8). As with the two-second utterance from the results, the ASR model failed to distinguish phonetic differences between Cantonese and Taishanese. 

    The findings come from a limited reference set constituting a small segment of the entire corpus. In community-led documentation projects, manual transcription is prohibitively labour-intensive \cite{foley_building_2018}. Even small fully verified segments represent substantial investment. As a pilot study, our evaluation therefore prioritises ecological validity over scale. In brief, Whisper does meet the needs of oral historians if the purpose of ASR tools is to provide a first-pass transcription of audio recordings \cite{prudhommeaux_automatic_2021}. In the case of our corpus, this is limited to the English-language segments and resourcing should be put towards correcting and editing the non-English segments. The results from this paper suggests that there is an unmet need to develop a reference set that reflects the linguistic situation of Cantonese-speaking heritage language communities.

\section{Conclusion}

    The primary contribution of this paper is that we show how existing state-of-the-art language technologies can be used to support heritage language revitalisation. Using the language-unspecified \textsc{large-v3} Whisper model, we applied ASR across the entire corpus. The entire process took five hours to transcribe 12 hours and 42 minutes of audio end-to-end which was less than 1\% of the estimated manual transcription time based on raw computational run time (excluding time to verify the transcripts). The benefit of using ASR toolkits such as Whisper to transcribe audio recordings is clear; however, it is important to account for the additional time needed to correct and edit the first-pass outputs. Future work should consider how optimisation processes such as fine-tuning using existing data can be used to improve the performance of code-switched multilingual speech. Of course, this step will necessitate engagement and consultation with the speakers and the wider community.

\section{Acknowledgements}

    We would like to thank the five community elders who generously provided their time to the Cantonese Heritage and Culture in Aotearoa New Zealand oral history project. This research was made possible through the Ngā Kōrero Tuku Iho, the New Zealand Oral History Grants, from Manatū Taonga | Ministry for Culture and Heritage. We would also like to acknowledge the support of the New Zealand Chinese Association (NZCA) Auckland Branch.

\section{Use of Generative AI Disclosure}

    The authors acknowledges the use of generative AI (Microsoft Copilot, GPT-5) for editing and polishing the manuscript following peer review to improve clarity, grammar, and consistency. No generative AI tools were used to produce substantive content.

\bibliographystyle{IEEEtran}
\bibliography{references}

@incollection{cheang_diglossia_2022,
	address = {Cham},
	title = {Diglossia in {Chinese}? {It}’s {Complicated}},
	isbn = {978-3-030-80072-7},
	doi = {10.1007/978-3-030-80072-7_7},
	booktitle = {Handbook of {Literacy} in {Diglossia} and in {Dialectal} {Contexts}: {Psycholinguistic}, {Neurolinguistic}, and {Educational} {Perspectives}},
	publisher = {Springer International Publishing},
	author = {Cheang, Leo Man-Lit and McBride, Catherine},
	editor = {Saiegh-Haddad, Elinor and Laks, Lior and McBride, Catherine},
	year = {2022},
	pages = {123--133},
}

@inproceedings{zhao_adapting_2025,
	title = {Adapting {Whisper} for {Code}-{Switching} through {Encoding} {Refining} and {Language}-{Aware} {Decoding}},
	doi = {10.1109/ICASSP49660.2025.10889634},
	booktitle = {Proceedings in 2025 {IEEE} {International} {Conference} on {Acoustics}, {Speech} and {Signal} {Processing}},
	author = {Zhao, Jiahui and Shi, Hao and Cui, Chenrui and Wang, Tianrui and Liu, Hexin and Ni, Zhaoheng and Ye, Lingxuan and Wang, Longbiao},
	month = apr,
	year = {2025},
	pages = {1--5},
}

@misc{stats_nz_2023_2024,
	title = {2023 {Census} place summaries},
	url = {https://tools.summaries.stats.govt.nz/},
	author = {{Stats NZ}},
	month = nov,
	year = {2024},
}

@inproceedings{vaswani_attention_2017,
	address = {Red Hook, NY},
	title = {Attention is all you need},
	booktitle = {Proceedings of the 31st {International} {Conference} on {Neural} {Information} {Processing} {Systems}},
	publisher = {Curran Associates},
	author = {Vaswani, Ashish and Shazeer, Noam and Parmar, Niki and Uszkoreit, Jakob and Jones, Llion and Gomez, Aidan N. and Kaiser, Lukasz and Polosukhin, Illia},
	month = dec,
	year = {2017},
	note = {https://dl.acm.org/doi/10.5555/3295222.3295349},
	pages = {6000--6010},
}

@book{norman_chinese_1988,
	address = {Cambridge, United Kindgom},
	title = {Chinese},
	isbn = {978-0-521-29653-3},
	publisher = {Cambridge University Press},
	author = {Norman, Jerry},
	year = {1988},
}

@inproceedings{draxler_transcribing_2025,
	title = {Transcribing {Oral} {History} {Recordings} {Using} the {Transcription} {Portal}},
	url = {https://www.isca-archive.org/interspeech_2025/draxler25_interspeech.html},
	booktitle = {Proceedings of the {Annual} {Conference} of the {International} {Speech} {Communication} {Association}},
	author = {Draxler, Christoph and Pömp, Julian and van den Heuvel, Henk and Ardolino, Fabio and van Hessen, Arjan},
	year = {2025},
	pages = {300--301},
}

@inproceedings{calamai_voices_2021,
	address = {Online},
	title = {Voices from {Ravensbrück}. {Towards} the creation of an oral and multilingual resource family},
	url = {https://usiena-air.unisi.it/handle/11365/1179271},
	booktitle = {Proceedings of {CLARIN} {Annual} {Conference} 2021},
	publisher = {Common Language Resources and Technology Infrastructure-European Research Infrastructure Consortium},
	author = {Calamai, Silvia and Beeken, J. and Van Den Heuvel, H. and Broekhuizen, M. and Van Hessen, A. and Draxler, Chr and Scagliola, S.},
	year = {2021},
}

@inproceedings{an_cantonese_2025,
	address = {Singapore, Singapore},
	title = {Cantonese {Dialect} {Transcription} in {Diverse} {Sophisticated} {Scenarios} via the {OpenAI} {Whisper} {Speech} {Recognition} {Model}},
	doi = {10.1007/978-981-96-7008-6_23},
	booktitle = {Neural {Information} {Processing}},
	publisher = {Springer Nature},
	author = {An, Jing and Bai, Yanbing and Li, Jiyi and Wang, Lifei and Jiang, Yuyi and Zhang, Yikui},
	editor = {Mahmud, Mufti and Doborjeh, Maryam and Wong, Kevin and Leung, Andrew Chi Sing and Doborjeh, Zohreh and Tanveer, M.},
	month = jul,
	year = {2025},
	pages = {317--328},
}

@article{chen_chinese_2018,
	title = {Chinese {Heritage} {Language} {Maintenance} in the {Context} of {Superdiversity}: {Perspectives} from {Dialect}-background {Heritage} {Learners}},
	volume = {4},
	doi = {10.1558/rtcfl.26170},
	number = {1},
	journal = {Researching and Teaching Chinese as a Foreign Language},
	author = {Chen, Lin and Wang, Danping},
	year = {2018},
	pages = {97--117},
}

@incollection{huang_overview_2010,
	title = {An {Overview} of {Modern} {Speech} {Recognition}},
	url = {https://www.microsoft.com/en-us/research/publication/an-overview-of-modern-speech-recognition/},
	booktitle = {Handbook of {Natural} {Language} {Processing}, {Second} {Edition}, {Chapter} 15 ({ISBN}: 1420085921)},
	author = {Huang, Xuedong and Deng, Li},
	year = {2010},
	pages = {339--366},
}

@misc{manatu_taonga__ministry_for_culture_and_heritage_processing_2024,
	title = {Processing the interview},
	url = {https://nzhistory.govt.nz/hands/processing-the-interview-a-guide-to-recording-oral-history},
	journal = {Oral history guide},
	author = {{Manatū Taonga {\textbar} Ministry for Culture and Heritage}},
	year = {2024},
}

@book{ritchie_doing_2015,
	address = {New York, NY},
	edition = {3},
	title = {Doing {Oral} {History}},
	publisher = {Oxford University Press},
	author = {Ritchie, Donald A.},
	year = {2015},
}

@incollection{ip_new_2005,
	address = {Wellington, New Zealand},
	title = {New {Zealand} {Chinese} identity: {Sojourners}, model minority and multiple identities},
	isbn = {978-0-86473-517-1},
	booktitle = {New {Zealand} {Identities}: {Departures} and {Destinations}},
	publisher = {Victoria University Press},
	author = {Ip, Manying and Pang, David},
	editor = {Liu, James Hou-fu and McCreanor, Tim and McIntosh, Tracey and Teaiwa, Teresia},
	year = {2005},
	pages = {174--190},
}

@incollection{ip_chinese_2013,
	address = {London, United Kingdom},
	title = {Chinese immigration to {Australia} and {New} {Zealand}: {Government} policies and race relations},
	doi = {10.4324/9780203100387},
	booktitle = {Routledge {Handbook} of the {Chinese} {Diaspora}},
	publisher = {Routledge},
	author = {Ip, Manying},
	editor = {Tan, Chee-Beng},
	year = {2013},
	pages = {156--175},
}

@book{bergen_transcribing_2019,
	address = {New York, NY},
	title = {Transcribing {Oral} {History}},
	publisher = {Routledge},
	author = {Bergen, Teresa},
	year = {2019},
}

@article{zhang_lora-int8_2025,
	title = {{LoRA}-{INT8} {Whisper}: {A} {Low}-{Cost} {Cantonese} {Speech} {Recognition} {Framework} for {Edge} {Devices}},
	volume = {25},
	issn = {1424-8220},
	doi = {10.3390/s25175404},
	number = {17},
	journal = {Sensors},
	author = {Zhang, Lusheng and Wu, Shie and Wang, Zhongxun},
	year = {2025},
	pages = {5404},
}

@incollection{ramesh_role_2025,
	address = {New York, NY},
	title = {The {Role} of {Oral} {Traditions} in {Language} {Preservation}},
	isbn = {979-8-3373-3730-2},
	doi = {10.4018/979-8-3373-3730-2.ch004},
	booktitle = {Preserving, {Documenting}, and {Revitalizing} {Surviving} {Dialects} and {Endangered} {Local} {Languages}},
	publisher = {IGI Global Scientific Publishing},
	author = {Ramesh, M. R.},
	editor = {Jomaa, Nayef J. and Al-Kathiri, Amir Azad Adli},
	year = {2025},
	pages = {93--138},
}

@incollection{yee_coping_2003,
	address = {Auckland, New Zealand},
	title = {Coping with {Insecurity}: {Everyday} {Experiences} of {Chinese} {New} {Zealanders}},
	isbn = {1-86940-289-8},
	booktitle = {Unfolding {History}, {Evolving} {Identity}: {The} {Chinese} in {New} {Zealand}},
	publisher = {Auckland University Press},
	author = {Yee, Beven},
	editor = {Ip, Manying},
	year = {2003},
	pages = {49--51},
}

@inproceedings{draxler_speech_2024,
	address = {Torino, Italy},
	title = {Speech {Technology} {Services} for {Oral} {History} {Research}},
	url = {https://aclanthology.org/2024.htres-1.6/},
	booktitle = {Proceedings of the {First} {Workshop} on {Holocaust} {Testimonies} as {Language} {Resources} ({HTRes}) @ {LREC}-{COLING} 2024},
	author = {Draxler, Christoph and van den Heuvel, Henk and van Hessen, Arjan and Ircing, Pavel and Lehečka, Jan},
	editor = {Anuradha, Isuri and Wynne, Martin and Frontini, Francesca and Plum, Alistair},
	month = may,
	year = {2024},
	pages = {38--43},
}

@inproceedings{james_developing_2020,
	address = {Cham, Switzerland},
	title = {Developing {Resources} for {Te} {Reo} {Māori} {Text} {To} {Speech} {Synthesis} {System}},
	doi = {10.1007/978-3-030-58323-1_32},
	booktitle = {Text, {Speech}, and {Dialogue}},
	publisher = {Springer International Publishing},
	author = {James, Jesin and Shields, Isabella and Berriman, Rebekah and Keegan, Peter J. and Watson, Catherine I.},
	editor = {Sojka, Petr and Kopeček, Ivan and Pala, Karel and Horák, Aleš},
	year = {2020},
	pages = {294--302},
}

@inproceedings{povey_kaldi_2011,
	address = {Prague, Czech Republic},
	title = {The {Kaldi} {Speech} {Recognition} {Toolkit}},
	url = {https://infoscience.epfl.ch/handle/20.500.14299/98397},
	booktitle = {{IEEE} 2011 {Workshop} on {Automatic} {Speech} {Recognition} and {Understanding}},
	author = {Povey, Daniel and Ghoshal, Arnab and Boulianne, Gilles and Burget, Lukas and Glembek, Ondrej and Goel, Nagendra and Hannemann, Mirko and Motlicek, Petr and Qian, Yanmin and Schwarz, Petr and Silovsky, Jan and Stemmer, Georg and Vesely, Karel},
	month = dec,
	year = {2011},
}

@inproceedings{radford_robust_2023,
	address = {Honolulu, HI},
	title = {Robust {Speech} {Recognition} via {Large}-{Scale} {Weak} {Supervision}},
	url = {https://proceedings.mlr.press/v202/radford23a.html},
	booktitle = {Proceedings of the 40th {International} {Conference} on {Machine} {Learning}},
	author = {Radford, Alec and Kim, Jong Wook and Xu, Tao and Brockman, Greg and Mcleavey, Christine and Sutskever, Ilya},
	month = jul,
	year = {2023},
	pages = {28492--28518},
}

@article{chen_micro_2024,
	title = {Micro language planning in {Mandarin}-dominated {Chinese} language education: voices from dialect-background heritage learners in {New} {Zealand}},
	volume = {25},
	doi = {10.1080/14664208.2023.2260634},
	number = {2},
	journal = {Current Issues in Language Planning},
	author = {Chen, Lin and Wang, Danping},
	year = {2024},
	pages = {157--175},
}

@inproceedings{foley_building_2018,
	title = {Building {Speech} {Recognition} {Systems} for {Language} {Documentation}: {The} {CoEDL} {Endangered} {Language} {Pipeline} and {Inference} {System} ({ELPIS})},
	doi = {10.21437/SLTU.2018},
	booktitle = {Proceedings of {The} 6th {International} {Workshop} on {Spoken} {Language} {Technologies} for {Under}-{Resourced} {Languages}},
	author = {Foley, B. and Arnold, J. and Coto-Solano, R. and Durantin, G. and Mark, E. and van Esch, D. and Heath, S. and Kratochvíl, F. and Maxwell-Smith, Z. and Nash, D. and Olsson, O. and Richards, M. and San, N. and Stoakes, H. and Thieberger, N. and Wiles, J.},
	month = aug,
	year = {2018},
}

@article{holmes_language_1993,
	title = {Language {Maintenance} and {Shift} in {Three} {New} {Zealand} {Speech} {Communities}},
	volume = {14},
	doi = {10.1093/applin/14.1.1},
	number = {1},
	journal = {Applied Linguistics},
	author = {Holmes, Janet and Roberts, Mary and Verivaki, Maria and Aipolo, Anahina},
	year = {1993},
	pages = {1--24},
}

@article{lynd_oral_1993,
	title = {Oral {History} from below},
	volume = {21},
	url = {https://www.jstor.org/stable/3675042},
	number = {1},
	journal = {The Oral History Review},
	author = {Lynd, Staughton},
	year = {1993},
	pages = {1--8},
}

@article{prudhommeaux_automatic_2021,
	title = {Automatic {Speech} {Recognition} for {Supporting} {Endangered} {Language} {Documentation}},
	volume = {15},
	url = {http://hdl.handle.net/10125/74666},
	journal = {Language Documentation \& Conservation},
	publisher = {University of Hawaii Press},
	author = {Prud'hommeaux, Emily and Jimerson, Robbie and Hatcher, Richard and Michelson, Karin},
	year = {2021},
	pages = {491--513},
}

@misc{gee_guangdong_2012,
	type = {Audio},
	title = {From {Guangdong} to {Aotearoa} - {An} oral history project},
	url = {https://natlib.govt.nz/records/44043481},
	author = {Gee, Suzanne},
	year = {2012},
}

@misc{chang_oral_2005,
	type = {Audio},
	title = {Oral history of {Chinese} women in {New} {Zealand}},
	url = {https://natlib.govt.nz/records/35853389},
	author = {Chang, Kitty},
	year = {2005},
}

@misc{tung_jung_association_of_new_zealand_inc_tung_2003,
	type = {Audio},
	title = {Tung {Jung} oral history project},
	url = {https://natlib.govt.nz/records/35850030},
	author = {{Tung Jung Association of New Zealand Inc} and Chang, Kitty},
	year = {2003},
}

@misc{thorpe_otaki_2004,
	type = {Audio},
	title = {Otaki {District} {Commercial} {Gardeners} {Society} oral history project},
	url = {https://natlib.govt.nz/records/35853904},
	author = {Thorpe, Agnes Anne and Bisdee, Margaret and {Otaki District Commercial Gardeners Society}},
	year = {2004},
}

@incollection{wang_chinese_2023,
	address = {Cham, Switzerland},
	title = {Chinese as a {Heritage} {Language} in {New} {Zealand}: {A} {Historical} {Overview}},
	isbn = {978-3-031-35475-5},
	shorttitle = {Chinese as a {Heritage} {Language} in {New} {Zealand}},
	url = {https://doi.org/10.1007/978-3-031-35475-5_2},
	doi = {10.1007/978-3-031-35475-5_2},
	language = {en},
	urldate = {2024-04-12},
	booktitle = {Teaching {Chinese} in the {Anglophone} {World}: {Perspectives} from {New} {Zealand}},
	publisher = {Springer International Publishing},
	author = {Wang, Danping},
	editor = {Wang, Danping and East, Martin},
	year = {2023},
	pages = {21--40},
}

@mastersthesis{ng_ninety_1962,
	address = {Christchurch, New Zealand},
	title = {Ninety {Years} of {Chinese} {Settlement} in {New} {Zealand}, 1866 to 1956},
	url = {http://dx.doi.org/10.26021/12112},
	language = {en},
	school = {University of Canterbury},
	author = {Ng, David},
	year = {1962},
}

\end{document}